\documentclass[11pt,twoside]{article}
\usepackage{amssymb}

\def\r{\mathbb R}                   

\begin{document}

\pagestyle{myheadings} \markboth {{\sc SUSY Classical Mechanics }}
{{\sc A. Alonso, M.A. Gonz\'alez and J.M. Guilarte }} \vspace{1cm}

%
%

\thispagestyle{empty}

\begin{center}

{\Large{\bf{Invariants in Supersymmetric}}}

\medskip

{\Large{\bf{Classical Mechanics}}}

\vskip1cm

{\sc A. Alonso Izquierdo$^1$, M.A.
Gonz\'alez Le\'on$^1$} \\
{\sc and J. Mateos Guilarte$^2$}
\vskip0.5cm

{\it $^1$ Departamento de Matem\'atica Aplicada. Universidad de Salamanca, SPAIN}
\vskip0.3cm
{\it $^2$ Departamento de
F\'{\i}sica. Universidad de Salamanca, SPAIN   }
\end{center}
\bigskip

%
%

\begin{abstract}
The bosonic second invariant of SuperLiouville models in
supersymmetric classical mechanics is described.
\end{abstract}

\section{Introduction.}

The search for kinks in N-component $(1+1)$-dimensional field
theory is a very difficult endeavour, see Reference \cite{Raja}
pp. 23-24. Mathematically, this problem is equivalent to a
N-dimensional mechanical system via dimensional reduction to
(0+1)-dimensions of space-time. Therefore, only if the mechanical
system is completely integrable is there a hope of finding all the
kinks of the field theory that are in one-to-one correspondence
with the separatrix trajectories (developing finite action in
infinite time) of the mechanical system. The authors have
investigated this line of research on one-dimensional topological
defects in several earlier papers, see \cite{Pa1}.

After the seminal paper of Olive and Witten on extended
supersymmetry algebras and solitons, \cite{Oliv}, the paradigmatic
N=1 kinks have been understood as BPS states of ${\cal N}=1$
supersymmetric quantum field theory. In $(1+1)$-dimensions, the
N=1 theory is based in an extended supersymmetric algebra because
there exist Majorana-Weyl fermions that could be used to
generalize the purely bosonic theory supporting kinks to a
Bose-Fermi system enjoying supersymmetry. Very recently, domain
walls have been studied in the Wess-Zumino model, \cite{Gibb}, and
${\cal N}=1$ SUSY QCD, \cite{Vali}. These theories arise in the
low energy limit of string theory and the domain walls can be seen
as membranes or other extended structures. On the other hand, both
the WZ model and SQCD contain more than one real scalar field.
Thus, their dimensional reduction is a $(1+1)$-dimensional either
${\cal N}=1$ or ${\cal N}=2$ supersymmetric field theory. Because
the kinks are the actual domain walls if they are looked at from a
$(3+1)$D point of view, in order to identify the BPS kinks (hence
the domain walls, hence the d-branes) it is convenient to study
${\cal N}=2$ or ${\cal N}=4$ supersymmetric classical mechanics,
starting from the finite action trajectories of these mechanical
systems.

Accordingly, in this work we shall study the ${\cal N}=2$
supersymmetric extension of the N=2 Liouville systems. These
supersymmetric extensions of completely integrable mechanical
systems arise in the dimensional reduction of $(1+1)$-dimensional
bosonic theories that have a very rich manifold of kinks. Our aim
is to elucidate wheter these supersymmetric models in classical
mechanics have a second invariant; if the answer is affirmative,
then in priciple all the finite action trajectories that give the
kinks and their supersymmetric partners can be found.

\section{Supersymmetric Classical Mechanics}

Let us start with a Grassman algebra $B_L$ with generators
$\theta_a^i \in B_L$, $a=1,2$, $i=1, \ldots ,N$,  that satisfy the
anti-commutation rules:
\[
\theta_a^i \theta_b^j + \theta_b^j \theta_a^i=0
\]
Any element of $B_L$ is a combination of the $L=2N$
generators $\theta_a^i$ that can be written in the form:
\[
b=b_0 {\bf 1}+\sum b_{i_1...i_m}^{a_1...a_m}
\theta_{a_1}^{i_1}...\theta_{a_m}^{i_m},
\]
where the coefficients $b_{i_1...i_m}^{a_1...a_m}$ are real
numbers. We shall distinguish between odd and even elements of
$B_L$ according to the parity of the number of Grassman
generators.

\subsection{Lagrangian Formalism}

The configuration space of the system is ${\cal C}={\r}^N \times
(B_{2N})$. If we choose $(x^i,\theta_1^i,\theta_2^i)$ as local
coordinates in ${\cal C}$, the Lagrangian of our system has the
\lq\lq supernatural" form:
\[
L=\displaystyle\frac{1}{2} \dot{x}^j \dot{x}^j
-U(\vec{x})+\displaystyle\frac{i}{2} \theta^j_a \dot{\theta}^j_a+
i R_{jk}(\vec{x}) \theta^j_1 \theta^k_2
\]
Observe that the Lagrangian is defined on even elements of ${\cal
C}$ of two types: there is a bosonic contribution collecting the
kinetic and potential energies of the $x^i$ coordinates. There are
also two terms that contain fermionic variables: besides the
Grassman kinetic energy, there is a Yukawa coupling $R_{jk}$ that
we assume to be symmetric in the $j,k$ indices, see \cite{Ca}. The
Euler-Lagrange equations
\[
\ddot{x}^k+\frac{\partial U}{\partial x^k}-i \frac{\partial
R_{jl}}{\partial x^k} \theta_1^j \theta_2^l=0 \hspace{1cm}
\dot{\theta}_1^i=-R_{ij} \theta_2^j \hspace{1cm}
\dot{\theta}_2^i=R_{ji} \theta_1^j
\]
determine the dynamics of our mechanical system. Via Noether's
theorem, one finds the energy as the invariant associated to
invariance under time translations:
\begin{equation}
I=H=\frac{1}{2} \dot{x}^j \dot{x}^j +U(\vec{x})- i R_{jk}(\vec{x})
\theta^j_1 \theta^k_2
\label{eq:hapseudo}
\end{equation}

\subsection{Hamiltonian Formalism}

We shall now briefly discuss the Hamiltonian formalism \cite{Ju}.
The usual definition of generalized momentum is extended to the
Grassman variables:
\begin{equation}
\pi_{\theta_a^j}= L
\frac{\stackrel{\leftarrow}{\partial}}{\partial \dot{\theta}_a^j}=
\frac{i}{2} \theta_a^j
\label{eq:momen}
\end{equation}
We note the dependence of the fermionic generalized momenta on
their Grassman variables. In the $6N$-dimensional phase space
$T^*{\cal C}$ with coordinates $(x^i,\theta_1^i,\theta_2^i,p_i,
\pi_{\theta_1^i},\pi_{\theta_2^i})$, the definition of Grassman
generalized momenta (\ref{eq:momen}) provides $2N$ second class
constraints. Instead of working in the reduced $4N$-dimensional
phase space (the Grassman sub-space of the phase space coincides
with the Grassman sub-space of the configuration space because the
Lagrangian is first-order in the velocities) we implement the
constraints in the Hamiltonian by means of Grassman-Lagrange
multipliers \cite{Di}:
\[
H_T=\displaystyle\frac{1}{2} \dot{x}^j \dot{x}^j +U(\vec{x})- i
R_{jk}(\vec{x}) \theta^j_1
\theta^k_2-(\pi_{\theta_a^j}-\displaystyle\frac{i}{2}
\theta_a^j) \lambda_a^j
\]
The Hamilton equations are:
\[
\dot{x}^j=\displaystyle\frac{\partial H_T}{\partial p_j}, \hspace{1cm} \dot{p}_j=-\displaystyle\frac{\partial H_T}{\partial x^j}, \hspace{1cm}  \dot{\theta}_a^j=\displaystyle\frac{\partial
H_T}{\partial \displaystyle\pi_{\theta_a^j}},
\hspace{1cm}
\dot{\pi}_{\theta_a^j}=\displaystyle\frac{\partial H_T}{\partial
\theta_a^j}
\]
Note the difference in sign between the bosonic and fermionic
canonical equations. Solving for the Lagrange multipliers, we find
\[
H_T=\frac{1}{2} \dot{x}^j \dot{x}^j +U(\vec{x})+ R_{jk}(\vec{x})
(\pi_{\theta_2^k} \theta_1^j-\pi_{\theta_1^j} \theta_2^k) ,
\]
which provides the right Hamiltonian flow in the phase space.

Defining the Poisson brackets for the generic functions $F$ and
$G$ in phase space in the usual way:
\begin{eqnarray*}
\{F,G\}_P & =& \frac{\partial F}{\partial p_j}\frac{\partial
G}{\partial q^j}-\frac{\partial F}{\partial q^j} \frac{\partial
G}{\partial p_j}+i F
\frac{\stackrel{\leftarrow}{\partial}}{\partial \theta_a^j}
\frac{\stackrel{\rightarrow}{\partial}}{\partial \theta_a^j}G-
\frac{1}{2} F \frac{\stackrel{\leftarrow}{\partial}}{\partial
\pi_{\theta_a^j}} \frac{\stackrel{\rightarrow}{\partial}}{\partial
\theta_a^j} G \\ & & -\frac{1}{2} F
\frac{\stackrel{\leftarrow}{\partial}}{\partial \theta_a^j}
\frac{\stackrel{\rightarrow}{\partial}}{\partial \pi_{\theta_a^j}}
G-\frac{i}{4} F \frac{\stackrel{\leftarrow}{\partial}}{\partial
\pi_{\theta_a^j}} \frac{\stackrel{\rightarrow}{\partial}}{\partial
\pi_{\theta_a^j}}  G ,
\end{eqnarray*}
the canonical equations read:
\[
\begin{array}{lcl}
\displaystyle\frac{d x^j}{d t}= \{H_T,x^j\}_P & \hspace{1cm} &
\displaystyle\frac{d \theta_a^j}{d t}= \{H_T,\theta_a^j\}_P
\\[0.3cm] \displaystyle\frac{d p_j}{d t}= \{H_T,p_j\}_P  & &
\displaystyle\frac{d \displaystyle\pi_{\theta_a^j}}{d t}=
\{H_T,\displaystyle\pi_{\theta_a^j}\}_P
\end{array}
\]
In general, the time-dependence of any observable $F$ is ruled by:
\[
\displaystyle\frac{d F}{d t}= \{H_T,F\}_P
\]
Therefore, physical observables are constants of motion or
invariants if and only if:
\[
\{H_T,I\}_P=0
\]

In practical terms, it is better to work on the reduced phase
space and define the reduced Poisson brackets as:
\[
\{F,G\}_P =\frac{\partial F}{\partial p_j}\frac{\partial
G}{\partial q^j}-\frac{\partial F}{\partial q^j} \frac{\partial
G}{\partial p_j}+i F
\frac{\stackrel{\leftarrow}{\partial}}{\partial \theta_a^j}
\frac{\stackrel{\rightarrow}{\partial}}{\partial \theta_a^j}G
\]
Thus, we receive the following Poisson structure:
\[
\{p_j,x^k\}_P=\delta_j^k \hspace{0.8cm} \{x^j,x^k\}_P=\{p_j,p_k\}_P=0 \hspace{0.8cm}
\{\theta_a^j, \theta_b^k\}_P=i \delta^{jk} \delta_{a b}
\]
and the canonical equations and the invariant observables are
referred to the reduced Hamiltonian $H$ given in (\ref{eq:hapseudo}).

\subsection{Supersymmetry}

The question arises: are there transformations in the
configuration or phase space such that they mix the bosonic and
fermionic variables and leave invariant the Lagrangian or the
Hamiltonian?. If the answer is yes, then the mechanical system can
be said to enjoy supersymmetry \cite{Fr}. Instead of using the
elegant superfield/superspace formalism, we take a direct
approach. Consider the following infinitesimal variations in the
configuration space defined in terms of a Grassman parameter
$\epsilon$:
\[
\begin{array}{lcl}
\begin{array}{c} \mbox{Variation 1:} \\
\left\{ \begin{array}{l} \delta_1 x^j=\varepsilon \theta_1^j \\
\delta_1 \theta_1^j= i \varepsilon \dot{x}^j \\ \delta_1
\theta_2^j=-i \varepsilon f^j(\vec{x}) \end{array} \right.
\end{array} & \hspace{2cm} & \begin{array}{c}
\mbox{Variation 2:} \\ \left\{ \begin{array}{l} \delta_2
x^j=\varepsilon \theta_2^j \\ \delta_2 \theta_1^j= i \varepsilon
g^j(\vec{x}) \\ \delta_2 \theta_2^j= i \varepsilon \dot{x}^j
\end{array} \right. \end{array}
\end{array}
\]
The variations induced on the Lagrangian are:
\begin{eqnarray*}
\delta_1 L&= & \displaystyle\frac{d}{dt}\left( \displaystyle\frac{1}{2} \dot{x}^j \varepsilon \theta_1^j \right) +\displaystyle\frac{1}{2} f^j \varepsilon \dot{\theta}_2^j- \left(R_{kj} \dot{x}^k+\displaystyle\frac{1}{2} \dot{f}^j \right) \varepsilon \theta_2^j- \\
&-& \left(\displaystyle\frac{\partial V}{\partial x^j}+ R_{jk} f^k \right) \epsilon \theta_1^j+
i \displaystyle\frac{\partial R_{jk}}{\partial x^l} \varepsilon \theta_1^l \theta_1^j \theta_2^k
\\
\delta_2 L&=& \displaystyle\frac{d}{dt}\left( \displaystyle\frac{1}{2} \dot{x}^j \varepsilon \theta_2^j \right) - \displaystyle\frac{1}{2} g^j \varepsilon \dot{\theta}_1^j+ \left(R_{kj} \dot{x}^k+\displaystyle\frac{1}{2} \dot{g}^j \right) \varepsilon \theta_1^j-\\ &-&\left(\displaystyle\frac{\partial V}{\partial x^j}+ R_{jk} g^k \right) \epsilon \theta_2^j+ i \displaystyle\frac{\partial R_{jk}}{\partial x^l} \varepsilon \theta_2^l \theta_1^j \theta_2^k
\end{eqnarray*}
There is symmetry with respect to the variations $\delta_1$ and
$\delta_2$ if and only if $\delta_1 L$ and $\delta_2 L$ are pure
divergences. This happens if
\[
U(\vec{x})=\displaystyle\frac{1}{2} \displaystyle\frac{\partial
W}{\partial x^j} \displaystyle\frac{\partial W}{\partial x^j}
\hspace{1.5cm} R_{jk}=-\frac{\partial^2 W}{\partial x^j \partial
x^k} \hspace{1.5cm} f^j=g^j=\displaystyle\frac{\partial
W}{\partial x^j} ,
\]
where $W(\vec{x})$ is a function defined in the bosonic piece of
the configuration space called the superpotential. A
supersymmetric Lagrangian has the form,
\[
L=\displaystyle\frac{1}{2} \dot{x}^j \dot{x}^j
+\displaystyle\frac{i}{2} \theta_a^j \dot{\theta}_a^j
-\displaystyle\frac{1}{2} \displaystyle\frac{\partial W}{\partial
x^j}\displaystyle\frac{\partial W}{\partial x^j}-i
\displaystyle\frac{\partial^2 W}{\partial x^j \partial x^k}
\theta_1^j \theta_2^k
\]
The Noether charges associated with the symmetries with respect to
$\delta_1$ and $\delta_2$ are respectively
\[
Q_1=\dot{x}^j \theta_1^j- \displaystyle\frac{\partial W}{\partial
x^j} \theta_2^j \hspace{2cm} Q_2=\dot{x}^j \theta_2^j+
\displaystyle\frac{\partial W}{\partial x^j} \theta_1^j
\]
Going to Hamiltonian formalism, one easily checks that these
fermionic supercharges close the  $N=2$ superalgebra
\[
\{Q_1,Q_1 \}_P=2 H \hspace{1.2cm} \{Q_1,Q_2\}_P=0 \hspace{1.2cm} \{Q_2,Q_2\}_P=2 H
\]
We immediately find a bosonic invariant, the Hamiltonian $H$
itself, and two fermionic constants of motion- the supercharges
$Q_1$ and $Q_2$; their Poisson brackets with $H$ are zero. If the
number of bosonic degrees of freedom is $N$, it is easy to check
that there is other bosonic invariant,
$I_3=\prod_{i=1}^N\theta_1^i \theta_2^i$ \cite{Pl}.

If the bosonic piece of the configuration space is a general
Riemannian manifold $M^N$ equipped with a metric tensor $g_{ij}$, the Grassman variables $\vartheta_a^i$ are the components of a
contravariant vector. The definition of the supercharges is generalized in the form:
\[
Q_1=g_{jk} \dot{x}^j \vartheta_1^k-\displaystyle\frac{\partial W}{\partial x^j} \vartheta_2^j \hspace{2cm}
Q_2=g_{jk} \dot{x}^j \vartheta_2^k+\displaystyle\frac{\partial W}{\partial x^j}\vartheta_1^j
\]
The supersymmetric algebra dictates the form of the Hamiltonian
\[
H=\frac{1}{2} g_{jk} \dot{x}^j \dot{x}^k + \frac{1}{2} g^{jk}
\frac{\partial W}{\partial x^j}\frac{\partial W}{\partial x^k} + i
W_{j;k} \vartheta^j_1 \vartheta^k_2,
\]
if
\[
W_{j;k}=\frac{\partial^2 W}{\partial x^j\partial
x^k}-\Gamma_{jk}^l\frac{\partial W}{\partial x^l},
\]
and the inverse Legendre transformation leads to the Lagrangian
\begin{equation}
L=\displaystyle\frac{1}{2} g_{jk} \dot{x}^j \dot{x}^k
+\displaystyle\frac{i}{2} g_{jk} \vartheta_a^j D_t
\vartheta_a^k+\displaystyle\frac{1}{4} R_{jkln} \vartheta_1^j
\vartheta_2^l \vartheta_1^k \vartheta_2^n
-\displaystyle\frac{1}{2} g^{jk}\frac{\partial W}{\partial x^j}
\frac{W}{\partial x^k} -i W_{j;k} \vartheta_1^j \vartheta_2^k ,
\label{eq:gene}
\end{equation}
where the covariant derivative is defined as $D_t \vartheta^j_a =\dot{\vartheta}^j_a + \Gamma^j_{lk} \dot{x}^l \theta^k_a$.

\section{From Liouville to SuperLiouville Models}

We shall focus on mechanical systems of two bosonic degrees of
freedom; $N=2$. In particular, we shall analyze the supersymmetric
extensions of the classical Liouville models. These models are
completely integrable and their classical invariants are well
known \cite{Li}. Our goal is to study what the invariants are in
their supersymmetric extensions.

In fact, the classical Liouville models are not only completely
integrable but Hamilton-Jacobi separable. The Hamilton-Jacobi
principle thus provides all the solutions of the dynamics in these
models. This is achieved by using appropriate coordinate systems.
There are four possibilities:

\vspace*{0.1cm}

\noindent $\bullet$ {\bf Liouville Models of Type I:} Let us consider the map $\xi^*:D
\longrightarrow {\r}^2$, where $D$ is an open sub-set of ${\r}^2$ with coordinates $(u,v)$. These variables are the elliptic
coordinates of the bosonic system if the map is defined as:
$\xi^{*}(x^1)=\frac{1}{\Omega} u v$, $\xi^{*}(x^2)=\frac{1}{\Omega} \sqrt{(u^2-\Omega^2)(\Omega^2-v^2)}$ and
 $u \in [\Omega,\infty)$, $v \in [-\Omega, \Omega]$.
In the new variables, the Lagrangian  of a Liouville model of Type
I reads:
\begin{equation}
L = \displaystyle\frac{1}{2}\, \displaystyle\frac{u^2-v^2}{u^2-\Omega^2} \, \dot{u}\, \dot{u}+ \displaystyle\frac{1}{2} \, \displaystyle\frac{u^2-v^2}{\Omega^2-v^2} \, \dot{v}\, \dot{v}
- \displaystyle\frac{u^2-\Omega^2}{u^2-v^2} \,  f(u)-\displaystyle\frac{\Omega^2-v^2}{u^2-v^2} \, g(v)
\label{eq:Liou1}
\end{equation}
Observe that apart from a common factor the contribution to the
Lagrangian of the $u$ and $v$ variables splits completely. The
common factor can be interpreted as a metric (of zero curvature):
$g_{ij}=(u^2-v^2) \delta_{ij}$.

\vspace*{0.1cm}

\noindent $\bullet$ {\bf Liouville Models of Type II:} In polar coordinates $\zeta^{*}(x^1)=\rho \cos \chi$, $\zeta^{*}(x^2)=\rho \sin \chi$, $\rho \in [0,\infty)$, $\chi\in [0,2\pi)$
the Lagrangian of the Liouville models of Type II reads:
\begin{equation}
L = \displaystyle\frac{1}{2} \, \dot{\rho} \, \dot{\rho} +
\displaystyle\frac{1}{2} \,\rho^2 \,\dot{\chi} \, \dot{\chi}
-f(\rho)-\frac{1}{\rho^2} \, g(\chi)
\label{eq:liou2}
\end{equation}
Again, besides the metric factor  $g_{11}=1$, $g_{22}=\rho^2$,
the contributions of $\rho$ and $\chi$ appear completely separated
in the Lagrangian.

\vspace*{0.1cm}

\noindent $\bullet$ {\bf Liouville Models of Type III:} Parabolic
coordinates $u\in (-\infty,\infty)$, $v\in [0,\infty)$ are defined
through the map $\gamma^{*}: D\longrightarrow {\r}^2$ such that
$\gamma^{*}(x^1)=\frac{1}{2} (u^2-v^2)$, $\gamma^{*}(x^2)=u v$. A
Liouville model of Type III obeys a Lagrangian of the form:
\begin{equation}
L = \displaystyle\frac{1}{2} (u^2+v^2) \left( \dot{u} \, \dot{u} +\dot{v} \, \dot{v} \right)
-\displaystyle\frac{1}{u^2+v^2} \left( f(u)+g(v) \right)
\label{eq:liou3}
\end{equation}
There is a metric factor $g_{ij}=(u^2+v^2) \delta_{ij}$ and
separate contributions of $u$ and $v$ to $L$.

\vspace*{0.1cm}

\noindent $\bullet$ {\bf Liouville Models of Type IV:} In these models the
Lagrangian is directly separated in Cartesian coordinates
\begin{equation}
L = \displaystyle\frac{1}{2} \,\dot{x}^1 \, \dot{x}^1 \,+\, \displaystyle\frac{1}{2} \, \dot{x}^2 \, \dot{x}^2\, -\,f(x^1)\,-\,g(x^2)
\label{eq:liou4}
\end{equation}
and a  Euclidean metric $g_{ij}=\delta_{ij}$ can be understood.

\vspace{0.1cm}

The definition of SuperLiouville models is a two step process:

$(i)$ Define a supersymmetric N=2 Lagrangian system on a
Riemannian manifold $M^2$, which is $\r^2$ equipped with the
metric induced by the maps $\chi^*$, $\zeta^*$ and $\gamma^*$ for
Types I, II, and III, and the Euclidean metric for Type IV.
Consider also Grassman variables that transform as
$\vartheta^i_a=\frac{\partial {x'}^i}{\partial x^j} \theta^j_a$
under these changes of coordinates.

$(ii)$ A model defined in this way is a SuperLiouville model if the
superpotential splits in such a manner that the bosonic part of
the Lagrangian coincides with the Lagrangian of a Liouville model.

\vspace*{0.1cm}

\noindent $\bullet$ {\bf  SuperLiouville Models of Type I:} The
Lagrangian of the bosonic sector of this Type of model contains
two contributions:
\[
L_B =\displaystyle\frac{1}{2}
\displaystyle\frac{u^2-v^2}{u^2-\Omega^2} \, \dot{u} \, \dot{u} +
\displaystyle\frac{1}{2} \displaystyle\frac{u^2-v^2}{\Omega^2-v^2}
\, \dot{v} \, \dot{v}- \displaystyle\frac{1}{2}
\displaystyle\frac{u^2-\Omega^2}{u^2-v^2} \left(
\displaystyle\frac{\partial W}{\partial u}
\right)^2-\displaystyle\frac{1}{2}
\displaystyle\frac{\Omega^2-v^2}{u^2-v^2} \left(
\displaystyle\frac{\partial W}{\partial v} \right)^2 ,
 \]
where the superpotential $W$ provides the potential $U_B$ through
the identity,
\[
U_{B}= \displaystyle\frac{1}{2} \displaystyle\frac{u^2-\Omega^2}{u^2-v^2} \left( \displaystyle\frac{\partial W}{\partial u} \right)^2+\displaystyle\frac{1}{2} \displaystyle\frac{\Omega^2-v^2}{u^2-v^2} \left( \displaystyle\frac{\partial W}{\partial v} \right)^2
\]
In the fermionic sector the Lagrangian takes the form
$L_F=T_F+L_{BF}^I$, with
\[
T_F=\frac{i}{2} \frac{u^2-v^2}{u^2-\Omega^2}
\vartheta_a^u D_t \vartheta_a^u+ \frac{i}{2}
\frac{u^2-v^2}{\Omega^2-v^2} \vartheta_a^v D_t \vartheta_a^v
\]
The Bose-Fermi interaction adds to the Lagrangian the Yukawa terms
\begin{eqnarray*}
L_{FB}^I & =& -i \left[ \frac{\partial^2 W}{\partial u \partial u}+\displaystyle\frac{\Omega^2-v^2}{(u^2-v^2)(u^2-\Omega^2)}\left(u \frac{\partial W}{\partial u} -v \frac{\partial W}{\partial v} \right)  \right] \vartheta_1^u \vartheta_2^u- \\ & & -i \left[ \frac{\partial^2 W}{\partial u \partial v} +\displaystyle\frac{1}{u^2-v^2} \left( v \frac{\partial W}{\partial u}-u \frac{\partial W}{\partial v} \right) \right] (\vartheta_1^u \vartheta_2^v+\vartheta_1^v \vartheta_2^u) \\ & &  -i \left[ \frac{\partial^2 W}{\partial v \partial v} +\displaystyle\frac{(u^2-\Omega^2)}{(u^2-v^2)(\Omega^2-v^2)} \left(u \frac{\partial W}{\partial u} -v \frac{\partial W}{\partial v} \right) \right] \vartheta_1^v \vartheta_2^v
\end{eqnarray*}

{\bf Definition:} A system in supersymmetric classical mechanics
is a SuperLiouville model of Type I if the map given by the change
from Cartesian to elliptic coordinates acting on the Cartesian
superpotential is such that:
\[
\xi^{*} W_C=W_1(u)\pm W_2(v)
\]

\vspace*{0.1cm}

\noindent $\bullet$ {\bf SuperLiouville Models of Type II:} The
bosonic Lagrangian is:
\[
L_B= \displaystyle\frac{1}{2}
\dot{\rho} \dot{\rho} + \frac{1}{2} \rho^2  \dot\chi \dot\chi -
\displaystyle\frac{1}{2} \left( \displaystyle\frac{\partial
W}{\partial \rho} \right)^2 -\frac{1}{2 \rho^2} \left(
\displaystyle\frac{\partial W}{\partial \chi} \right)^2
\]
and the superpotential $W$ is related to the potential through
\[
U_{B}= \displaystyle\frac{1}{2} \left( \displaystyle\frac{\partial
W}{\partial \rho} \right)^2+ \frac{1}{2 \rho^2}\left(
\displaystyle\frac{\partial W}{\partial \chi} \right)^2
\]
In the fermionic sector, the Lagrangian takes the form:
\[
T_F=\frac{i}{2} \vartheta_a^{\rho} D_t \vartheta_a^{\rho}+
\frac{i}{2} \rho^2 \vartheta_a^\chi  D_t \vartheta_a^\chi ,
\]
and the Yukawa Bose-Fermi couplings are:
\begin{eqnarray*}
{L}_{FB}^I&=& -i \frac{\partial^2 W}{\partial \rho \partial \rho} \vartheta_1^\rho \vartheta_2^\rho- i \left( \frac{\partial^2 W}{\partial \chi \partial \chi} +\rho \frac{\partial W}{\partial \rho} \right) \vartheta_1^\chi\vartheta_2^\chi-\\
& &  -i \left( \frac{\partial^2 W}{\partial \rho \partial \chi} -\displaystyle\frac{1}{\rho} \frac{\partial W}{\partial \chi} \right) (\vartheta_1^\rho \vartheta_2^\chi+\vartheta_1^\chi
\vartheta_2^\rho )
\end{eqnarray*}

{\bf Definition:} a system in supersymmetric classical mechanics
is a SuperLiouville Model of Type II if the map given by the
change from Cartesian to polar coordinates acting on the Cartesian
superpotential produces:
\[
\varsigma^{*} W_C= W_1(\rho)\pm W_2(\chi)
\]

\vspace*{0.1cm}

\noindent $\bullet$ {\bf SuperLiouville Models of Type III:} The
bosonic Lagrangian is:
\[
L_B= \displaystyle\frac{1}{2}
(u^2+v^2) \left( \dot{u} \dot{u}+ \dot{v} \dot{v}  \right) -
\displaystyle\frac{1}{2 (u^2+v^2)} \left[ \left(
\displaystyle\frac{\partial W}{\partial u} \right)^2+ \left(
\displaystyle\frac{\partial W}{\partial v} \right)^2 \right]
\]
The potential is determined from the superpotential $W$ as:
\[
U_{B}= \displaystyle\frac{1}{2 (u^2+v^2)} \left[ \left( \displaystyle\frac{\partial W}{\partial u} \right)^2+
\left( \displaystyle\frac{\partial W}{\partial v} \right)^2 \right]
\]
The purely fermionic contribution to the Lagrangian is:
\[
T_F=\frac{i}{2} (u^2+v^2) \vartheta_a^u D_t
\vartheta_a^u+ \frac{i}{2} (u^2+v^2) \vartheta_a^v D_t \vartheta_a^v
\]
and the Yukawa couplings are:
\begin{eqnarray*}
{L}_{BF}^I&=& -i \left[ \frac{\partial^2 W}{\partial u \partial u} -\displaystyle\frac{1}{u^2+v^2} \left(u \frac{\partial W}{\partial u} -v \frac{\partial W}{\partial v} \right) \right] \vartheta_1^u \vartheta_2^u- \\ & & -i \left[ \frac{\partial^2 W}{\partial u \partial v} - \displaystyle\frac{1}{u^2+v^2} \left( u \frac{\partial W}{\partial v} +v \frac{\partial W}{\partial u} \right) \right] (\vartheta_1^u \vartheta_2^v+\vartheta_1^v \vartheta_2^u) \\
& & -i \left[ \frac{\partial^2 W}{\partial v \partial v} +\displaystyle\frac{1}{u^2+v^2} \left( u \frac{\partial W}{\partial u} -v \frac{\partial W}{\partial v} \right) \right] \vartheta_1^v \vartheta_2^v
\end{eqnarray*}

{\bf Definition:} A system in supersymmetric classical mechanics
is a  SuperLiouville Model of Type III if the map given by the
change from Cartesian to parabolic coordinates acting on the
Cartesian superpotential is such that:
\[
\rho^{*} W_C= W_1(u)\pm W_2(v)
\]

\vspace*{0.1cm}

\noindent $\bullet$ {\bf SuperLiouville Models of Type IV:}
Finally, the definition of SuperLiouville Model of Type IV is
straightforward.

{\bf Definition:} A system in supersymmetric classical mechanics
belongs to Type IV SuperLiouville models if the superpotential
$W(x^1,x^2)$ is of the form:
\[
W(x^1,x^2)=W_1(x^1)\pm W_2(x^2)
\]
The Lagrangian is:
\[
L = \displaystyle\frac{1}{2} \dot{x}^j \dot{x}^j
+\displaystyle\frac{i}{2} \theta_a^j \dot{\theta}_a^j
-\displaystyle\frac{1}{2} \displaystyle\frac{\partial W}{\partial
x^j} \displaystyle\frac{\partial W}{\partial x^j}-i
\displaystyle\frac{\partial^2 W}{\partial x^1 \partial x^1}
\theta_1^1 \theta_2^1-i \displaystyle\frac{\partial^2 W}{\partial
x^2 \partial x^2} \theta_1^2 \theta_2^2 ,
\]
and the system can be be understood as a ${\cal N}=2 \oplus {\cal
N}=2$ SUSY in $(0+1)$ dimensions.

\vspace*{0.1cm}

As a common feature, observe that the potential is insensitive to
the relative signs of the separated parts of the superpotential.
Therefore, all the Liouville models are supersymmetrizable by
means of two different superpotentials.

\section{On the Bosonic Invariants}
It is well known that Liouville models have a second invariant in
involution with the energy -the first invariant- that guarantees
complete integrability in the sense of the Liouville theorem. We
shall now show that SuperLiouville models also have a second
invariant of bosonic nature. Our strategy in the search of such an
invariant, $[I,H]=0$, follows the general pattern shown in the
literature: see \cite{H1}. The ansatz for invariants at the
highest quadratic level in the momenta is:
\[ \begin{array}{lcr}
I&=&\displaystyle{\frac{1}{2} H^{ij} p_i p_j + V(x_1,x_2)+F_{ij}
\theta_1^i \theta_2^j+G_{ij} \theta_1^i \theta_1^j+ J_{ij}
\theta_2^i \theta_2^j+} \\ & & \displaystyle{+ L^i_{jk} p_i
\theta_1^j \theta_1^k+ M^i_{jk} p_i \theta_2^j \theta_2^k+
N^i_{jk} p_i \theta_1^j \theta_2^k+ S_{ijkl} \theta_1^i \theta_2^k
\theta_1^j \theta_2^l}
\end{array}
\]
Here, we assume that:

\vspace*{0.1cm}

\noindent - $i)$ $H^{ij}$ is a symmetric tensor depending on $x^i$. There are
three independent functions to determine.

\vspace*{0.1cm}

\noindent - $ii)$ $L^i_{jk}$ and $M^i_{jk}$ also depend only on
$x^i$ and are antisymmetric in the indices $j$ and $k$:
$L^i_{jk}=-L^i_{kj}$, $M^i_{jk}=-M^i_{kj}$. They include four
independent functions.

\vspace*{0.1cm}

\noindent - $iii)$ $G_{ij}$ and $J_{ij}$ are antisymmetric
functions of $x^i$ in the indices: $G_{ij}=-G_{ji}$ and
$J_{ij}=-J_{ji}$. $F_{ij}(x^i)$, however, is neither symmetric nor
antisymmetric; it contains four independent functions.

\vspace*{0.1cm}

\noindent - $iv)$ Finally, $S_{ijkl}(x^i)$ is antisymmetric in the exchange of the indices $i,j$ and $k,l$ and symmetric in the exchange of the pairs $ij,kl$. There is only one independent function to determine in this tensor.

\vspace*{0.1cm}

\noindent The commutator with the Hamiltonian is:
\begin{eqnarray*}
& & \hspace{-0.6cm}  -[I,H] = {\displaystyle  \frac{1}{2} \frac{\partial H^{jk}}{\partial x^l} p_l
p_j p_k +\left( -H^{lj} \frac{\partial^2 W}{\partial
x^j\partial x^k} \frac{\partial W}{\partial x_k}+\frac{\partial
V}{\partial x_l} \right) p_l +}
\\
& & \hspace{-0.6cm} {\displaystyle + \left(-i H^{nj}\frac{\partial^3 W}{\partial x^k\partial x^l\partial x^j}+ \frac{\partial F_{kl}}{\partial x_i}+2
L_{km}^i\frac{\partial^2 W}{\partial x_m\partial x_l}+2  M_{ln}^i
\frac{\partial^2 W}{\partial x_n\partial x^k} \right) p_n
\theta_1^k \theta_2^l+} \\
& & \hspace{-0.6cm} {\displaystyle  + \left(\frac{\partial^2
W}{\partial x_l \partial x^k} F_{lj}- M^n_{kj}
\frac{\partial^2 W}{\partial x_n\partial x_l}\frac{\partial
W}{\partial x^l} \right) \theta_2^k \theta_2^j +\left( \frac{\partial J_{lj}}{\partial
x_k}+N^k_{mj} \frac{\partial^2 W}{\partial x_m\partial x^l}
\right) p_k \theta_2^l \theta_2^j-}\\
& & \hspace{-0.6cm} {\displaystyle -
\left(\frac{\partial^2 W}{\partial x^j\partial x_k}
F_{nk}+ L^l_{nj} \frac{\partial^2 W }{\partial
x^l\partial x^k}\frac{\partial W}{\partial x_k} \right) \theta_1^n
\theta_1^j+ \left( \frac{\partial G_{nj}}{\partial
x_l}-N^l_{nk}\frac{\partial^2 W}{\partial x_j\partial x_k} \right)
p_l \theta_1^n \theta_1^j+}
\\
& & \hspace{-0.6cm} {\displaystyle  +\, 2 \left( G_{nj}\frac{\partial^2 W}{\partial x_j\partial x^k}+ J_{kl}\frac{\partial^2 W}{\partial x^n\partial x_l}-\frac{1}{2}
N^j_{nk} \frac{\partial^2 W}{\partial x^j\partial
x^l}\frac{\partial W}{\partial x_l} \right) \theta_1^n \theta_2^k+}
\\
& & \hspace{-0.6cm} {\displaystyle + \frac{\partial L^n_{jk}}{\partial x_l} p_l p_n
\theta_1^j \theta_1^k+ \frac{\partial M^n_{jk}}{\partial x_l} p_l
p_n \theta_2^j \theta_2^k+ \frac{\partial N^n_{jk}}{\partial x_l}
p_l p_n \theta_1^j \theta_2^k+} \\
& & \hspace{-0.6cm} {\displaystyle -i N^n_{jk}\frac{\partial^3 W}{\partial x^n\partial x^l\partial
x^m} \theta_1^j \theta_2^k \theta_1^l \theta_2^m -i \frac{\partial
S_{njkl}}{\partial x_m} p_m \theta_1^n \theta_2^k \theta_1^j
\theta_2^l }
\end{eqnarray*}
\noindent Therefore, $[I,H]=0$, and $I$ is a second invariant if
and only if the following equations are satisfied:

\begin{center}
\begin{tabular}{|l|l|}
\hline &
\\
{\small BOX 1} & \mbox{\bf a)}
$\displaystyle\frac{\partial H^{ij}}{\partial x^k}+\frac{\partial
H^{kj}}{\partial x^i}=0$ \\ & \\
\hline & \\ {\small BOX 2} & \mbox{\bf a)} $\displaystyle H^{ij} \frac{\partial^2 W}{\partial x^j\partial x^k}\frac{\partial W}{\partial x_k}=\frac{\partial V}{\partial x_i}$  \\ & \\
\hline & \\  {\small BOX 3} & \mbox{\bf a)} $\displaystyle{\epsilon^{jk} \frac{\partial
L^i_{jk}}{\partial x_l}+\epsilon^{jk} \frac{\partial
L^l_{jk}}{\partial x_i} =0 }$ \\ & \mbox{\bf b)}
$ \displaystyle{\epsilon^{jk} \frac{\partial M^i_{jk}}{\partial x_l}+\epsilon^{jk}
\frac{\partial M^l_{jk}}{\partial x_i} =0}$   \\ & \\
\hline &\\  {\small BOX 4} & \mbox{\bf a)} $\displaystyle{H^{nj}\frac{\partial^3 W}{\partial x^k\partial x^l\partial x^j}+i \frac{\partial F_{kl}}{\partial x_n}+2 i
L_{km}^n\frac{\partial^2 W}{\partial x_m\partial x_l}+2 i M_{lj}^n
\frac{\partial^2 W}{\partial x_j\partial x^k}=0}$ \\ & \mbox{\bf b)} $\displaystyle{\epsilon^{ij}\left( \frac{\partial^2
W}{\partial x_i \partial x^k} F_{kj}- M^k_{ij}
\frac{\partial^2 W}{\partial x_k \partial x_l}\frac{\partial
W}{\partial x^l} \right) =0}$ \\ & \mbox{\bf c)} $\displaystyle{\epsilon^{ij} \left(\frac{\partial^2
W}{\partial x^j \partial x_k} F_{ik}+ L^l_{ij}
\frac{\partial^2 W}{\partial x^k\partial x^l}\frac{\partial
W}{\partial x_k} \right)=0}$ \\ & \\ \hline & \\
  {\small BOX 5} & \mbox{\bf a)} $\displaystyle\epsilon^{ij} \left( \frac{\partial G_{ij}}{\partial
x_l}-N^l_{jk}\frac{\partial^2 W}{\partial x_i \partial x_k}
\right)=0$
\\ &\mbox{\bf b)} $\displaystyle \epsilon^{ij} \left( \frac{\partial J_{ij}}{\partial
x_k}+N^k_{mj} \frac{\partial^2 W}{\partial x_m\partial x^i}
\right)=0 $ \\ &\mbox{\bf c)} $\displaystyle G_{ij}\frac{\partial^2 W}{\partial x_j\partial x^k}+
J_{kl}\frac{\partial^2 W}{\partial x^i\partial x_l}-\frac{1}{2}
N^j_{ik} \frac{\partial^2 W}{\partial x^j\partial
x^l}\frac{\partial W}{\partial x_l} =0 $
\\ & \mbox{\bf d)} $\displaystyle \frac{\partial N^i_{jk}}{\partial x_l}+\frac{\partial N^l_{jk}}{\partial x_i}=0 $\\ & \mbox{\bf e)} $\displaystyle \epsilon^{ij} \epsilon^{lk}
N^m_{jk}\frac{\partial^3 W}{\partial x^i\partial x^l\partial
x^m}=0 $ \\ &\\ \hline &\\ {\small BOX 6} & \mbox{\bf a)} $\displaystyle \epsilon^{ij} \epsilon^{kl} \frac{\partial
S_ {ijkl}}{\partial x_m}=0$ \\ &\\ \hline
\end{tabular}
\end{center}

\subsection{General properties of the solution}
We deal with a overdetermined system of partial differential
equations: there are 31 PDE relating 15 unknown functions.
Moreover, some sub-systems can be solved for some sub-set of
functions. We proceed in a recurrent way:

\vspace*{0.1cm}

- $i)$ The equations in {\it BOXES 1} and {2} are sufficient to
find  $H^{ij}$ and $V$. We recover the information about the
second invariant of the purely bosonic sector: the Liouville
model.

\vspace*{0.1cm}

- $ii)$ The equations in {\it BOX 3} are solved if the independent
components of $L^i_{jk}$ and $M^i_{jk}$ have the form,
\[
L^i_{12}=C\, \epsilon^{ij} x_j +A_i \hspace{2cm} M^i_{12}=D \,
\epsilon^{ij} x_j +B_i ,
\]
where $A_i$, $B_i$, $C$ y $D$ are constants.

\vspace*{0.1cm}

- $iii)$ The equations in {\it BOX 4}, together with the previous
information, leads to the  computation of $F_{ij}$. The identity
$\frac{\partial^2 F_{kl}}{\partial x_1\partial
x_2}=\frac{\partial^2 F_{kl}}{\partial x_2\partial x_1}$ and the
equation 4a) requires that
\[
\epsilon^{mn} \displaystyle\frac{\partial}{\partial x_m} \left[ L^n_{jk}
\frac{\partial^2 W}{\partial x_j\partial x^l}+
M^n_{jl}\frac{\partial^2 W}{\partial x^k\partial_j}+\frac{i}{2}
H^{nj}\frac{\partial^3 W}{\partial x^j\partial x^k\partial x^l}
\right]=0
\]
Moreover, if we restrict $F_{ij}$ to be symmetric under the
exchange of indices and then identify $L^i_{jk}=M^i_{jk}$,
equation 4b) becomes equal to 4c).

\vspace*{0.1cm}

- $iv)$ The equations of {\it BOX 5} are satisfied if :
\[
G_{ij}=J_{ij}=N_{ijk}=0
\]

\vspace*{0.1cm}

- $v)$ Equation 6a) by itself, {\it BOX 6}, sets the only
independent component of  $S_{ijkl}$ to be constant; $
S_{1212}=\mbox{cte}$. Then:
\[
I_3=\theta_1^1 \theta_1^2 \theta_2^1 \theta_2^2
\]
is a constant of motion, an invariant.

\subsection{Invariants in SuperLiouville models}

We now apply the previous results to the computation of the
supersymmetric extensions of the second invariant of Liouville
models. In general they have the form:
\[
I_2=I_2^{(B)}+I_2^{(F)},
\]
where $I_2^{(B)}$ is the \lq\lq body", already present in the Liouville
model, and $I_2^{(F)}$ is the \lq\lq soul"-containing Grassman
variables- of the second invariant in the SuperLiouville models.
We find:

\subsubsection{SuperLiouville Models of Type I:}
\begin{eqnarray*}
I_2^{(B)}& =&\frac{1}{2} \left[ \left(x^2 \dot{x}^1- x^1 \dot{x}^2
\right)^2- \Omega^2 \dot{x}^2 \dot{x}^2 + \left(x^2 \frac{\partial
W}{\partial x^1} - x^1 \frac{\partial W}{\partial x^2}\right)^2-
\Omega^2 \frac{\partial W}{\partial x^2}\frac{\partial W}{\partial
x^2} \right] \\ I_2^{(F)}&=&i (x^2 \dot{x}^1 - x^1 \dot{x}^2)
\theta_a^1 \theta_a^2+i\left(2 x^1 \frac{\partial^2 W}{\partial
x^2
\partial x^2}-\frac{\partial W}{\partial x^1}-x^2 \frac{\partial^2
W}{\partial x^1 \partial x^2} \right) \theta_1^2 \theta_2^2 + \\ &
+&i \, \left(-x^1 x^2 \frac{\partial^2 W}{\partial x^2 \partial
x^2}+x^2 \frac{\partial W}{\partial x^1}+ x^2 x^2 \frac{\partial^2
W}{\partial x^1 \partial x^2}\right) (\theta_1^1
\theta_2^2+\theta_1^2 \theta_2^1) +\\ &  +&i \, \left(- x^2
\frac{\partial W}{\partial x^2}- x^1 x^2 \frac{\partial^2
W}{\partial x^1 \partial x^2}+ x^2 x^2 \frac{\partial^2
W}{\partial x^1 \partial x^1} \right) \theta_1^1 \theta_2^1
\end{eqnarray*}

\subsubsection{SuperLiouville Models of Type II:}
\begin{eqnarray*}
I_2^{(B)}& = & \frac{1}{2} \left(x^2 \dot{x}^1 - x^1 \dot{x}^2 \right)^2+ \frac{1}{2} \left(x^2 \frac{\partial W}{\partial x^1} - x^1 \frac{\partial W}{\partial x^2} \right)^2
\\
I_2^{(F)}&=& i \left( x^2 \dot{x}^1 - x^1 \dot{x}^2 \right)
\theta_a^1 \theta_a^2+i x^2 \left(x^2 \frac{\partial W}{\partial
x^1 \partial x^1}-x^1 \frac{\partial^2 W}{\partial x^1
\partial x^2}-\frac{\partial W}{\partial x^2}\right) \theta_1^1
\theta_2^1+ \\ & +& i x^2\left(x^2 \frac{\partial^2 W}{\partial
x^1 \partial x^1}+\frac{\partial W}{\partial x^1}-x^1
\frac{\partial^2 W}{\partial x^2 \partial x^2}\right)(\theta_1^1
\theta_2^2+\theta_1^2 \theta_2^1)+ \\ &+& i x^1 \left(x^1
\frac{\partial^2 W}{\partial x^2  \partial x^2}-\frac{\partial
W}{\partial x^1}-x^2 \frac{\partial^2 W}{\partial x^1 \partial
x^2}\right) \theta_1^2 \theta_2^2
\end{eqnarray*}

\subsubsection{SuperLiouville Models of Type III:}
\begin{eqnarray*}
I_2^{(B)}&=&\left(x^1 \dot{x}^2 - x^2 \dot{x}^1 \right) \dot{x}^2+
\left(x^1 \frac{\partial W}{\partial x^2} - x^2 \frac{\partial
W}{\partial x^1} \right)\frac{\partial W}{\partial x^2} \\
I_2^{(F)}&=&- i \dot{x}^2 \theta_a^1 \theta_a^2- i x^2
\frac{\partial^2 W}{\partial x^1 \partial x^2} \theta_1^1
\theta_2^1-i x^2 \frac{\partial^2 W}{\partial x^2 \partial x^2}
(\theta_1^1 \theta_2^2+\theta_1^2 \theta_2^1) +\\ & &+i\left(2 x^1
\frac{\partial^2 W}{\partial x^2 \partial x^2}-\frac{\partial
W}{\partial x^1}-x^2 \frac{\partial^2 W}{\partial x^1 \partial
x^2}\right) \theta_1^2 \theta_2^2
\end{eqnarray*}

\subsubsection{SuperLiouville Models of Type IV:}
\[
I_2^{(B)}= \frac{1}{2} \dot{x}^1 \dot{x}^1 +\frac{1}{2} \frac{\partial W}{\partial x^1} \frac{\partial W}{\partial x^1}  \hspace{2cm}
I_2^{(F)}=  i \frac{\partial^2 W}{\partial x^1 \partial x^1} \theta_1^1 \theta_2^1
\]

Finally, we briefly comment on the geometrical and physical
meaning of the second invariant. Usually, it is related to
transformation that is termed as a hidden symmetry. We see that by
introducing the generalized momenta $\Pi_j={\dot
x}_j+i\frac{\partial W}{\partial x^j}$, the second invariant of
the Type I model is:
\[
I_2^{(B)}=\frac{1}{2} \left[ \left| x^2\Pi_1-x^1\Pi_2
\right|^2-\Omega^2 \left| \Pi_2 \right|^2 \right],
\]
which is no more than the modulus of the generalized angular
momentum to the square added to $-\Omega^2$ times the square of
$|\Pi_2|$. Similar considerations are easily applied to the second
invariant of the other Types. A generalized momentum such as
$\Pi_j$ can be obtained if one adds the complex topological piece:
\[
L_T^{(B)}=i{\dot x}^j\frac{\partial W}{\partial x^j}
\]
to the bosonic Lagrangian.

\end{document}